\documentclass[journal=jctcce,manuscript=article]{achemso}

\usepackage[T1]{fontenc} 
\usepackage{graphicx}
\usepackage{dcolumn}
\usepackage{bm}
\usepackage{textcomp}
\usepackage[usenames,dvipsnames]{xcolor}

\usepackage{amsmath} 
\usepackage{amssymb} 
\usepackage{amsfonts}

\usepackage{array}

\usepackage{multirow}
\usepackage{hyperref} 
\usepackage{romanbar}

\usepackage{appendix}

\author{Karol Palczynski}
\email{karol.palczynski@helmholtz-berlin.de}
\affiliation{Research Group Simulations of Energy Materials, Helmholtz-Zentrum Berlin f\"ur Materialien und Energie, Hahn-Meitner-Platz 1, D-14109 Berlin, Germany}

\author{Thorren Kirschbaum}
\email{thorren.kirschbaum@helmholtz-berlin.de}
\affiliation{Research Group Simulations of Energy Materials, Helmholtz-Zentrum Berlin f\"ur Materialien und Energie, Hahn-Meitner-Platz 1, D-14109 Berlin, Germany}
\alsoaffiliation{Institute of Mathematics, Freie Universit\"at Berlin, Arnimallee 14, D-14195 Berlin, Germany}

\author{Annika Bande}
\email{annika.bande@helmholtz-berlin.de}
\affiliation{Theory of Electron Dynamics and Spectroscopy, Helmholtz-Zentrum Berlin f\"ur Materialien und Energie, Hahn-Meitner-Platz 1, D-14109 Berlin, Germany}

\author{Joachim Dzubiella}
\email{joachim.dzubiella@physik.uni-freiburg.de}
\affiliation{Applied Theoretical Physics - Computational Physics, Albert-Ludwigs-Universit\"at Freiburg, Hermann-Herder Stra{\ss}e 3, D-79104 Freiburg, Germany}
\alsoaffiliation{Research Group Simulations of Energy Materials, Helmholtz-Zentrum Berlin f\"ur Materialien und Energie, Hahn-Meitner-Platz 1, D-14109 Berlin, Germany}

\title{Hydration structure of diamondoids from reactive force fields}

\begin{document}

\begin{abstract}
Diamondoids are promising materials for applications in catalysis and nanotechnology. 
Since many of their applications are in aqueous environments, to understand their function it is essential to know the structure and dynamics of the water molecules in their first hydration shells.
In this study, we develop an improved reactive force field (ReaxFF) parameter set for atomistically resolved molecular dynamics simulations of hydrated diamondoids to characterize their interfacial water structure.
We parameterize the force field and validate the water structure against geometry-optimized structures from density functional theory.
We compare the results to water structures around diamondoids with all partial charges set to zero, and around charged smooth spheres, and find qualitatively similar water structuring in all cases.
However, the response of the water molecules is most sensitive to the partial charges in the atomistically resolved diamondoids.  
From the systematic exclusion of atomistic detail we can draw generic conclusions about the nature of the hydrophobic effect at nanoparticle interfaces and link it to the interfacial water structure.
The interactions between discrete partial charges on short length scales affect the hydration structures strongly but the hydrophobic effect seems to be stable against these short scale surface perturbations.
Our methods and the workflow we present are transferable to other hydrocarbons and interfacial systems.
\end{abstract}

\section{Introduction}
Nanodiamonds (NDs) are nanometer-sized undoped, hydrogen terminated fragments of the diamond crystal lattice.
The smallest NDs form a subclass called diamondoids (DDs), also named polymantanes, which are constructed from the smallest unit cage structure of the diamond lattice, the molecule adamantane (AD) with the molecular formula C$_{10}$H$_{16}$.
DDs and NDs are an important class of hydrocarbons with applications in nanotechnology~\cite{Ge2014,Henych2019,Khajeh2019,Aranifard2020,Yeung2020,Miliaieva2021} and biomedicine~\cite{Mochalin2011,Lamoureux2010,G2013,Jariwala2020}, due to their physical stability, chemical inertness and high surface-to-volume ratio~\cite{Dahl2003,Stehlik_2016,Machova2020}.
For many applications, the molecules are solvated in aqueous media.
The interactions between the water molecules and the diamond-based solutes as well as the structure of the water molecules at the solute-water interfaces significantly affect the function of these materials~\cite{Horinek2008}.
However, the nanoscale hydration structures of hydrogenated NDs and DDs are still under debate. 

In experiments, the interfacial water molecules around NDs of various surface terminations have been probed using X-ray absorption~\cite{Petit_2015, Petit2017}, infrared~\cite{Ji_1998,Stehlik_2016,Petit2017} or Raman and photoluminescence~\cite{Dolenko_2012} spectroscopy, as well as differential scanning calorimetry~\cite{Stehlik_2016} or Fabry-Perot interferometry~\cite{Batsanov_2014}.
However, all these methods leave much room for interpretation as direct imaging of the interfacial water molecules is not yet possible~\cite{Galamba2013,Graziano2014}.
Other related studies shed some light on the water structure by combining experiments and computer simulations of water around hydrophobic solutes~\cite{Galamba2013}. 
Classical atomistic molecular dynamics (MD) computer simulations provide an idealized model of hydrated NDs and DDs.
The hydration properties of the solutes depend significantly on the electrostatic interactions, which in MD simulations are governed by the Coulomb potential, and so they depend on the partial charges on the atoms of the hydrated molecules~\cite{Maciel2012,https://doi.org/10.48550/arxiv.2102.13312, https://doi.org/10.48550/arxiv.2102.09187, https://doi.org/10.48550/arxiv.2103.01385}.
There are many different approaches implement partial charges in MD simulations.
A critical difference between all methods is that the charges on the atoms are either fixed, i.e. constant in time, or the atoms are polarizable, so that the charges respond to their environments. 
Both ways have been applied in MD simulations of NDs and DDs.
We will next review studies that used fixed partial charges, then experiments, and then simulations with polarizable charges.

Fixed partial charges are still the standard in MD simulations. 
For instance, fixed partial charges have been used to simulate the distribution and dynamics of functionalized AD molecules in a lipid bilayer~\cite{Chew2008}. 
Combined with classical force fields, the AD charges (with implicit hydrogens) were obtained from quantum density functional theory (DFT) calculations. 
Others have investigated the surface charge density from the zeta potential of an ND slab in water using a classical force field~\cite{Ge2016}. 
The partial charges were calculated with the so-called extended bond-charge increment scheme~\cite{Vanommeslaeghe2012}.
In yet other studies, default partial charges implemented in simulation software packages have been used to simulate binding of NDs to supramolecular structures~\cite{Schnbeck2018,Wang2020}.
Of particular importance for the present work are studies that report properties of hydrated AD molecules.
Two studies have simulated density distributions and orientations of water molecules around AD monomers~\cite{Ohisa2011} and dimers~\cite{Makowski2010}.
These studies found that the water molecules in the first hydration shells are mostly oriented tangential to the AD surface.
In another study, thermodynamic hydration properties for the AD molecule and bigger DDs have been calculated from MD~\cite{Maciel2012}. As most did before, the authors used classical force fields and the fixed partial charges were calculated with DFT. 
The MD simulations reveal that the enthalpy of hydration is negative, but it is compensated by the entropic contribution to the hydration free energy.
However, the balance between enthalpy and entropy sensitively depends on the force field. 
Some force fields with fixed non-zero charges yield negative hydration free energies, which is inconsistent with the low water solubilities of DDs observed in experiments~\cite{Maciel2012}, whereas, when all solute charges are set to zero, AD molecules are clearly hydrophobic~\cite{Bogunia_2022}.

Experiments confirm the hydrophobic character of DDs.
Measurements of DD solubilities in pressurized hot water showed that DDs are less soluble in water (i.e. more hydrophobic) than most aromatic hydrocarbons of the same carbon number~\cite{solubility_in_pressurized_hot_water_2008}. 
For instance, at 313~K and 500 atmospheres, the aqueous solubility of AD is 240 times lower than that of naphthalene. 
This effect can be emphasized by the tendency of the DDs to nucleate in water due to strong van der Waals interactions between them~\cite{Maciel2012}.
However, the authors of the experiments state that nucleation did not affect the resulting solubility values. 
Measurements of thermodynamic properties of host-guest complexation with DDs also agree that DDs are strongly hydrophobic~\cite{complexation_2005,complexation_2006,complexation_2017}. 
Interestingly, X-ray absorption and Raman measurements of small hydrogenated NDs~\cite{Petit2017} suggest that they may have a very unusual hydration structure as compared to other hydrocarbons:
Even though the hydrophobic nature of these NDs is not at all contested, they are actually disrupting the water hydrogen bond network surrounding them. 
A decrease of the OH bending vibrational mode reveals the presence of non-hydrogen bonded water in the first hydration shell with dangling OH groups from the water molecule.
These dangling OH groups seem to be associated with the hydrogenated ND surface groups.
Dangling OH groups in the hydration shells of hydrophobic solutes, including NDs, have been discussed in earlier studies~\cite{Mizuno2009,Perera2009,TomlinsonPhillips2011,Davis2012,Davis2013,Hlzl2018,https://doi.org/10.48550/arxiv.2102.09187}.
Close to the interface, a significant amount of water molecules points with their OH groups toward the hydrophobic interface, where the water molecules take on a slightly ice-like order~\cite{Hlzl2018}. 
The probability for the formation of dangling OH bonds depends, among others, on the solute's charge~\cite{Davis2013}.
Under certain conditions, i.e., if a solute contains polar groups, water molecules can also take part in weak C-H$\cdots$O-H$_2$ hydrogen bonds~\cite{Mizuno2009}, which can have a high influence on molecular conformations~\cite{Scheiner_2011}. 
Experimentally, however, C-H$\cdots$O-H$_2$ bonds are difficult to confirm and to distinguish from conventional water-water hydrogen bonds~\cite{Mizuno2009}.
Raman measurements of NDs show no traces of CH surface groups forming hydrogen bonds with water molecules~\cite{Petit2017}. 
Hydrogenated NDs and DDs are strongly hydrophobic, but quantum calculations show that the binding energy of a single water molecule to AD in vacuum is $0.92$~kcal$/$mol~\cite{Ohisa2011}, very weak compared to the water-water hydrogen bond strength of 5.5~kcal$/$mol in bulk-water at normal conditions~\cite{Kananenka2020}, but not negligible.

For classical force fields to be able to replicate the physics of hydrated DDs, MD simulations may require the use of force fields that explicitly address the effects of electrostatic polarization to incorporate a higher level of realism~\cite{Maciel2012}. 
One work analyzed the effects of ND additives on the adsorption of tricresyl phosphate on iron oxide surfaces~\cite{Khajeh2019}.
The authors used the ReaxFF force field with a parameter set optimized for interactions between hydrocarbons and metals in water~\cite{Reaxff2001,Aryanpour2010}.
The partial charges were calculated with a charge equilibration method (QEq)~\cite{QEq_Mortier1986,QEq_Rappe1991,QEq_Nakano1997,QEq_Aktulga2012}. 
A few studies applied the "Gasteiger" method, which is also related to electronegativity equalization, to calculate the partial charges of AD and bigger DDs for interactions with biopolymers in water~\cite{Luzhkov2012,Aranifard2020}.
There is still no consensus on a force field for the simulation of NDs and DDs, or in particular for the interactions between DDs and water. 
The hydration structure of NDs and DDs has not yet been studied with polarizable partial charges.
For that purpose, the QEq method seems promising, but it requires a carefully chosen set of parameters. 
Additionally, the ReaxFF force field will enable us to study reactive surface chemistry, such as reactions of graphitic carbon with water~\cite{Chang_2018,Mikheev_2020,Petit_2012}, solvated electrons~\cite{Zhang_2014,Brun_2020,Buchner_2021} (through eReaxFF e.g.~\cite{Islam2016}), and the influence of surface modifications~\cite{Gunawan_2014,Larsson_2018,Jariwala_2020,Kirschbaum2022}, in the future. 
For simulations of hydrocarbons in water, there are three notable QEq parameter sets for the ReaxFF force field published in literature.
The "Aryanpour" parameter set was first designed for interactions between hydrocarbons and metals~\cite{SanzNavarro2008} and later extended to work well in water~\cite{Aryanpour2010}. 
The "Wang" parameter set can describe chemical reactions of C- and Si-based solids with water, H$_{2}$, and O$_{2}$ molecules~\cite{Wang2020b}.
The "Zhang" parameter set focuses primarily on the correct bulk water structure and dynamics of water molecules in the condensed phase, but also considers the weak interactions between hydrocarbons and water~\cite{Zhang2018}.

The goal of this study is to find a ReaxFF force field that is able to reproduce the hydrophobic nature of hydrogenated DDs and to investigate the hydration structure of AD and the DDs diamantane (DI), triamantane (TR), and hexamantane (HE).
For this, we first calculate and characterize a realistic benchmark structure of the hydration shell around an AD molecule from DFT.
We then simulate the same system with the ReaxFF parameter sets of Aryanpour, Wang and Zhang and compare the results against the benchmarks.
Since it turns out that none of these force fields satisfactorily reproduces our references, we will adjust the Zhang parameter set to the DFT results by tuning its QEq parameters.
After that, we extrapolate the new force field to the other DDs and to room temperature in bulk-water at normal conditions to predict their hydration structure.

\section{Methods}

\subsection{DFT calculations}
\label{methods:dft_calculations}
\begin{figure}
\includegraphics[width=0.5\textwidth]{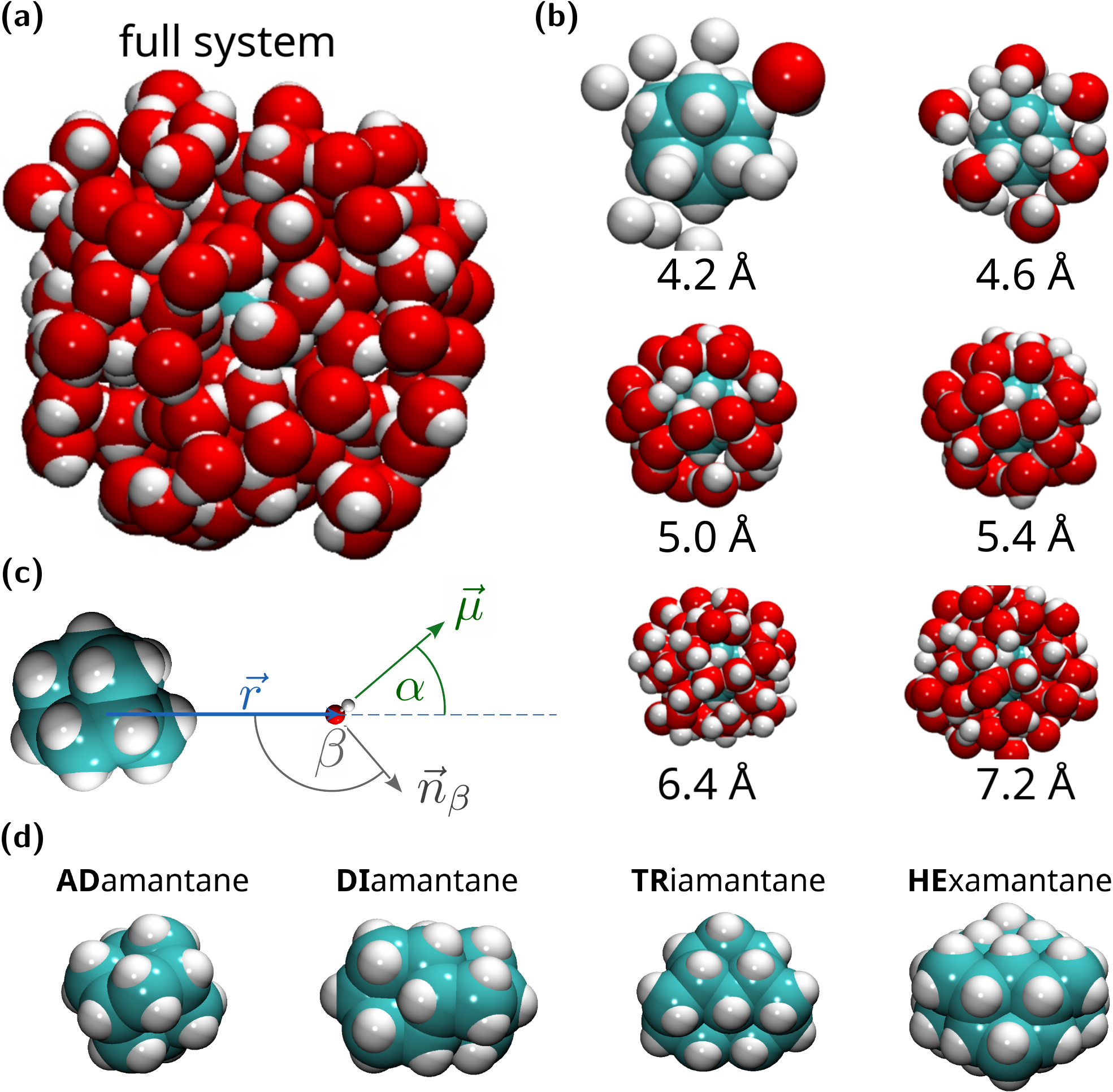}
\caption{
Illustrations of the simulated DD-water systems and important structural parameters.
a) The adamantane (AD) surrounded by a droplet of 143 water molecules, optimized with the revPBE functional and def2-SVP basis set with dispersion correction. 
b) To illustrate the hydration-shell structure, the smaller pictures show the same system but the water atoms are cut-off from view if their distance $\left|\vec{r}\right|$ to the AD center-of-mass (COM) is bigger than the displayed value.
c) Illustration of the angles and vectors that describe the position and orientation of a water molecule relative to the AD COM.
The angle $\alpha$ is defined between the water dipole moment vector $\vec{\mu}$ and the vector $\vec{r}$ connecting the AD COM to the water oxygen. 
The angle $\beta$ is defined between the normal of the water molecular plane, $\vec{n_\beta}$, and the vector $\vec{r}$.
e) MD snapshots of the diamondoids considered in this study: adamantane ("AD", C$_{10}$H$_{16}$), diamantane ("DI", C$_{14}$H$_{20}$), triamantane ("TR", C$_{18}$H$_{24}$) and hexamantane ("HE", C$_{26}$H$_{30}$).
}
\label{fig:illustrations}
\end{figure}
We optimize 21 random initial geometries of a single AD molecule (see Figure \ref{fig:illustrations}) surrounded by a shell of 143 water molecules (and also ten geometries of a single DI, TR and HE each surrounded by 160 water molecules) using the ORCA quantum chemistry program package~\cite{Neese2011,Neese2017} version 4.2.1.
We employ the revPBE functional~\cite{zhang_comment_1998, gillan_perspective_2016} and the def2-SVP basis set~\cite{schafer_fully_1992, weigend_balanced_2005} with dispersion correction~\cite{grimme_effect_2011} while using loose optimization criteria.
The functional is no hybrid functional and the basis set is relatively small, yet the lack of detail is a necessary compromise for optimizing bigger geometries and the method has proven itself before~\cite{gillan_perspective_2016, Ganji2017}. 
One geometry-optimized structure is depicted in Figure~\ref{fig:illustrations} both with the full water environment (panel a) and with only sublayers of the same (panel b).
The structures are then analyzed to identify properties that can be used as benchmarks to compare different MD force fields to.

\subsection{MD minimization with existing ReaxFF force fields}
\label{methods:md_simulations}
After establishing our DFT benchmarks, we proceed with MD simulations. 
We perform MD minimization~\cite{Bitzek2006} at $T=0$~K using the ReaxFF force field with the different parameter sets of Aryanpour~\cite{Aryanpour2010}, Wang~\cite{Wang2020b} and Zhang~\cite{Zhang2018}. 
The algorithm iteratively adjusts the atomic coordinates in such a way that, governed by the force field, all atoms move into the local potential energy minima that are closest to the initial structures. 
With each parameter set, we minimize all 21 previously geometry-optimized AD$+$water structures.
The simulations are computationally inexpensive, however, they do not exhaustively sample the phase space, so the results can not make any statement about the statistical probability of the configurations. 
Regardless, as long as the DFT structures are representative of stable states in the MD simulations, then the structures should change as little as possible during the minimization.
After that, we compare the MD-minimized structures to the DFT-optimized structures in terms of the benchmark properties. 

\subsection{Fitting of the force field parameters}\label{sec:fitting}
We modify the Zhang parameter set such that the MD-minimized structures achieve the optimal agreement with the DFT-optimized structures with respect to the benchmarks. 
We choose the Zhang set for its strength in regards to water simulations~\cite{Zhang2018}.
A full optimization of all ReaxFF force field parameters is, however, outside of the scope of this study. 
For the modifications we tune the parameters associated with the QEq charge equilibration for the carbon atoms of the AD, as we will explain later.
The QEq method is based on the electronegativity equalization principle which states that, at equilibrium, the electron density transfers between all atoms such that the electronegativities at all atomic sites are equal~\cite{QEq_Mortier1986}.
The combination of partial charges that fulfills this condition is found by minimizing the electrostatic energy of the system after every reasonable number of time steps. 
There are three QEq parameters: the shielding, the electronegativity and the hardness. 
We tune the shielding value in the range of 0.1 to 1~\AA$^{-1}$ with steps of 0.025~\AA$^{-1}$, the electronegativity value in the range of 1 to 12~eV with steps of 0.25~eV, and the hardness value in the range of 1 to 8~eV with steps of 1~eV.
For each parameter combination, we again minimize the previously DFT-optimized structures and calculate the averaged values for the benchmark properties. 
To find the optimal parameter set, we minimize the cost function

\begin{equation}
\delta x=\sum_{i}\left[\frac{x_{i}^{{\rm DFT}}-x_{i}^{{\rm ReaxFF}}}{\sigma_{i}}\right]^{2},\label{eq:cost_function}
\end{equation}

where $x_{i}^{{\rm DFT}}$ are the values of the benchmark quantities from the DFT calculations, and $x_{i}^{{\rm ReaxFF}}$ are the their counterparts from the MD simulations. 
The coefficient $\sigma_{i}$ gives each benchmark quantity a weight, necessary to balance the influence that each quantity has on the cost function.
This is an upper tolerance limit for the differences between the DFT and MD results. 
The weights are all about 10\% of the maximum value in which the corresponding benchmark quantities from the DFT-optimization range.
The parameter set which yields energy-minimized structures that minimize the cost function is next used to predict the water structure around the AD in bulk-water at room temperature ($T=300$~K).

\subsection{MD simulations at room temperature}
\label{subsec:AD_300K_sim}
Next, we prepare the simulations of an AD molecule solvated in bulk-water. 
We use a cubic simulation box with periodic boundary conditions.
We first fill the box with 343 water molecules and equilibrate them for 10~ps in the $NVT$-ensemble at $T=300$~K and at a particularly low density of $0.5$~g/cm$^3$ so that we can later insert an AD molecule into the box without its atoms overlapping with water molecules.  
We seperately pre-equilibrate the AD molecule in vacuum without periodic boundary conditions and then insert it into the water box so that the center-of-mass (COM) of the AD coincides with the center of the box. 
If the AD is still overlapping with water molecules, these molecules are removed from the simulation.
In the subsequent equilibration in the $NPT$-ensemble, the temperature and pressure are controlled by a Nose-Hoover thermostat and barostat. 
They are set to $T=300$~K and $P=1$~atm, while the damping constants are set to 25~fs and 250~fs, respectively.
The timestep is 1~fs.
Meanwhile, the AD is retained in the box center by subtracting the COM velocity of the AD from all atoms in the system. 
The equilibration is finished after 3~ns, the  density is constant and slightly above 1~g/cm$^3$, and the system's total energy is constant as well.
The production simulation runs for 1.5~ns with a time step of 0.5~fs.

\section{Results and discussion}

\subsection{DFT calculations}
\label{sec:DFT_calculations}
\begin{figure*}
\includegraphics[width=1\textwidth]{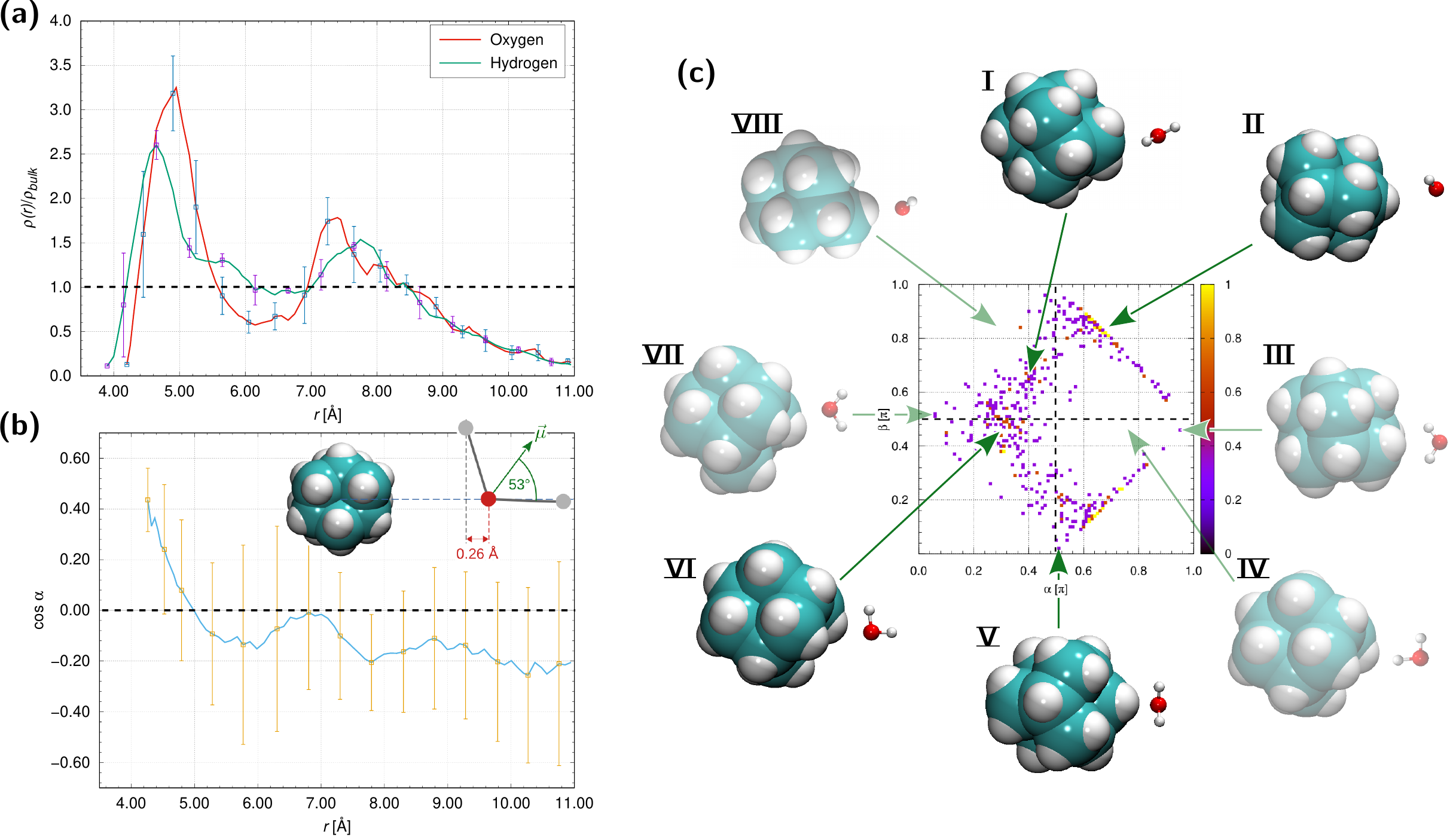}
\caption{
Results of the DFT geometry optimization of an AD surrounded by a droplet of 143 water molecules employing the revPBE functional and def2-SVP basis set with dispersion correction. 
a) The radial density of the water O and H atoms as function of the distance to the AD COM, normalized by the bulk density, 0.03344~\AA$^{-3}$ for O and 0.06688~\AA$^{-3}$ for H. 
b) The orientation of the water dipole moments as function of the distance to the AD COM. 
The error bars in (a) and (b) are the standard deviations from a set of 21 geometries after optimization.
The inset illustrates a water molecule with $\alpha=53$\textdegree, which is the average orientation closest to the AD. The illustration explains why the first peak of the H atom density is 0.26~\AA\ closer to the AD than that of the C atom density.
c) The relationship between the angles $\alpha$ and $\beta$ observed in the DFT-optimized structures in the first hydration shell. 
The color map indicates a relative probability for a molecule to occupy a certain $\left(\alpha,\beta\right)$ state. 
By dividing the map into sections, we can distinguish between eight types of similarly oriented water molecules which we call orientational modes. 
}
\label{fig:DFT_results}
\end{figure*}

We start by summarizing the results of the DFT geometry optimizations of the water droplets around the DDs.
As a basis for comparing the structures that we will obtain from MD simulations to the DFT-optimized structures, we define a set of benchmark quantities that characterize the alignment of the water molecules around the AD. 

The {first} quantity is the radial density distribution of the water O and H atoms as function of the distance $r$ to the AD COM (Figure \ref{fig:DFT_results} a). 
In the first hydration shell, which stretches from $4$ to $6.2$~\AA\ (as defined by the first minimum of the O density), the water H atoms tend to rest $0.3$~\AA\ closer to the AD than the O atoms (see also Figure \ref{fig:illustrations} b).
Above $6.2$~\AA, the region between the two peaks of the water density distribution forms a $1$~\AA\ wide hydrogen-dominated layer (see also Figure \ref{fig:illustrations} b).
Note that both densities go to zero at high $r$, because the optimized structures are droplets and not periodic bulk water.

The {second} quantity is the orientation angle $\alpha$ of the water dipole moments $\vec{\mu}$ as a function of the distance to the AD COM as defined in Figure~\ref{fig:illustrations}~c. 
Each dipole moment vector $\vec{\mu}$ is pointing from the O atom to the space halfway between the H atoms. 
The angle $\alpha$ is defined between $\vec{\mu}$ and the vector $\vec{r}$ pointing from the AD COM to the water O.
One might assume then from the relative positions of the H and O density peaks in Figure \ref{fig:DFT_results} (a) that most water dipole moments in the first hydration shell are pointing towards the AD. 
However, Figure \ref{fig:DFT_results} (b) reveals that the opposite is true.
Close to the AD, $\cos(\alpha)$ is positive, which means that $\vec{\mu}$ is pointing away from the AD. 
Yet, panels (a) and (b) do not contradict each other. 
Close to the AD, the average $\alpha$ value is 53\textdegree. 
As shown in the inset of Figure \ref{fig:DFT_results} (b), there is a range of geometries where $\alpha=53$\textdegree ~while one H atom is $\sim0.26$~\AA\ closer to the AD COM than the O atom. 
This explains the relative shift between the O and H peak in Figure \ref{fig:DFT_results} (a) despite $\vec{\mu}$ pointing away from the AD. 

For the {third} benchmark quantity, we additionally introduce the angle $\beta$ between the normal of the water molecular plane and $\vec{r}$ (see Figure \ref{fig:illustrations} c). 
In combination with the angle $\alpha$, we obtain a detailed description of the alignment of the water molecules in the first hydration shell. 
Figure \ref{fig:DFT_results} (c) shows the relationship between the angles $\alpha$ and $\beta$ found in the DFT-optimized structures. 
The map can be divided up in sections, with each section corresponding to similarly oriented water molecules.  
For instance, the molecules corresponding to the top and bottom edges of that map ($\alpha \approx 0.5\pi$, $\beta \approx \pm 1\pi$) are aligned parallel to the AD surface, with only small variations, and the dipole directions are oriented tangential to the AD surface.
The frequent occurrence of such oriented molecules is consistent with previous Monte Carlo and MD simulations that used non polarizable force fields~\cite{Ohisa2011,Doi2013}.
According to these studies, the water molecules in the first hydration shell are tangential to the skeleton surface of the AD. 
However, this is the single only orientational mode reported.
Our simulations yield a spectrum of eight distinct modes of orientation, which we each number with a roman numeral. 
Many molecules occupy a bulge-like area close to the left edge of the map. 
These molecules are oriented as illustrated in the inset of Figure \ref{fig:DFT_results} (b).
The other orientational modes are illustrated in Figure \ref{fig:DFT_results} (c).
We identify modes \Romanbar{1}, \Romanbar{2}, \Romanbar{5} and \Romanbar{6}, because they occur with a particularly high rate in the DFT results.
Conversely, the other $\alpha$-$\beta$ combinations (modes \Romanbar{3}, \Romanbar{4}, \Romanbar{7} and \Romanbar{8}) have a remarkably low occurrence on the map. 
The orientations of the water molecules are also related to the presence of dangling OH bonds in the first hydration shell, which is further discussed in appendix~\ref{app:dang}.
The $\alpha$-$\beta$ map is a fingerprint of the water structure in the first hydration shell and serves as the third benchmark for the upcoming MD simulations.

The {fourth} quantity is the net charge of the AD. 
The net charges of the AD in the DFT calculations are determined with the ESP fit method and range from -0.356~e to +0.025~e with an average of -0.183~e.
In accordance with previous investigations, there is a small electron transfer from water towards the AD~\cite{Petrini2007}.

The {fifth} quantity is the number of C-H$\cdots$O-H$_2$ dipole bonds (quasi hydrogen bonds) between the AD's H atoms and the water O atoms, where the C atoms are the donors and the water O atoms are the acceptors.
We define a non-covalent bond as a dipole bond if the distance between the acceptor and the donor is lower than 3.5~\AA \  and the C-H$\cdots$O bond angle is smaller than 30\textdegree. 
The time-averaged total number of AD-water dipole bonds in the DFT results is 1.3, i.e., only approx. $8$\% of the AD hydrogens are engaged in a dipole bond.

\subsection{Benchmarking the ReaxFF force fields and fitting to DFT}
\label{sec:Benchmarking_results}
\begin{figure*}
\includegraphics[width=1\textwidth]{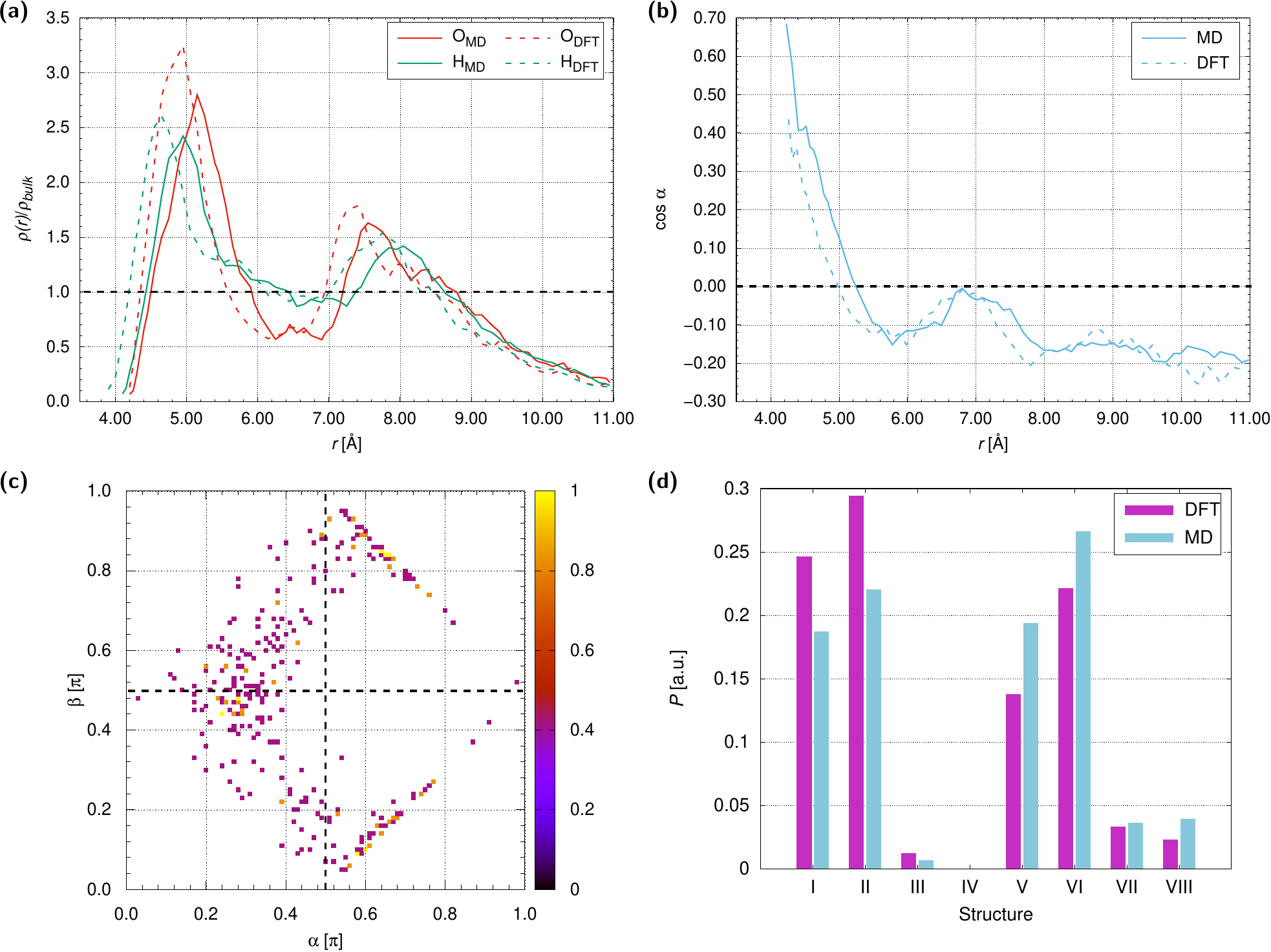}
\caption{
Summary of the 0~K minimization with the ReaxFF force field using the modified Zhang parameters.
Comparison between the MD results and the DFT benchmarks.
a) The radial density of the water O and H atoms as function of the distance to the AD COM, normalized by the bulk density, 0.03344~\AA$^{-3}$ for O and 0.06688~\AA$^{-3}$ for H.
b) The orientation of the water dipole moments as function of the distance to the AD COM.
c) The relationship between the angles $\alpha$ and $\beta$ in the first hydration shell obtained with the modified Zhang parameters.
d) The probabilities to find a water molecule in one of the structural modes which are defined in Figure \ref{fig:DFT_results} c.
}
\label{fig:MD_Zhang_0.225_7.00_min}
\end{figure*}
\begin{figure*}
\includegraphics[width=1\textwidth]{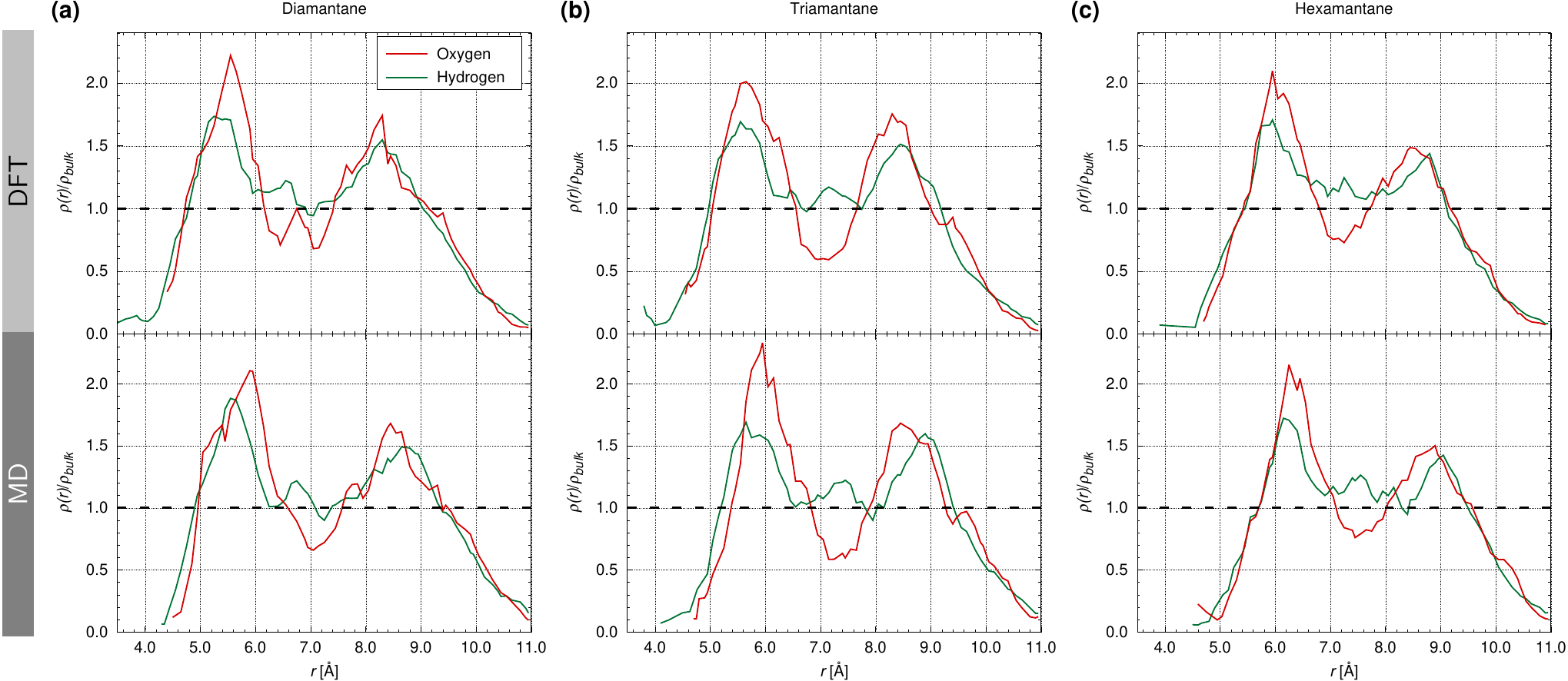}
\caption{
Radial density distributions of the water O and H atoms as function of the distance to the diamondoid COM, normalized by the bulk density, 0.03344~\AA$^{-3}$ for O and 0.06688~\AA$^{-3}$ for H.
The structures are comprised of a diamondoid surrounded by a water droplet of 160 molecules.
Comparison between DFT geometry optimization (top) and MD minimization with the modified Zhang parameters at $T=0$~K (bottom).
}
\label{fig:MD_Zhang_0.225_7.00_DI_TR_HE_water_density_distributions}
\end{figure*}

After the benchmarks have been established, we minimize the previously DFT-optimized structures using the ReaxFF force fields with the Aryanpour, Wang and Zhang parameter sets.
The deviations from the DFT-optimized structures are analyzed and presented in the appendix section \ref{appA}.

In short, we find that the commonly used ReaxFF parameters do not yield a good agreement with the DFT structures. 
The density distributions and orientations of the water molecules in respect to the AD COM are not correctly reproduced, the net charge of the AD is too high, and the number of AD-water dipole bonds is overestimated by a factor of $\sim 4$. 
Thus, to fit the ReaxFF simulations to the DFT-optimized structures, it is necessary to modify one of the existing parameter sets. 
The procedure of fitting the QEq parameters, that govern the charge equilibration of the C atoms, is discussed in the appendix section \ref{appB}.

We find that the cost function (equation~\ref{eq:cost_function}) is minimized by QEq parameters that compromise on all benchmark quantities.
\begin{center}
\begin{tabular}{rl}
shielding & $\gamma=0.225$~\AA\ \tabularnewline
electronegativity & $\chi=7.00$~eV \tabularnewline
hardness & $\eta=7.0601$~eV\tabularnewline
\end{tabular}
\end{center}
The electronegativity corresponds to 2.788 on the Pauling scale and is close to the value for an isolated C atom (2.55) and much closer than the default value from the original Zhang parameter set (1.819).
The net charges and number of AD-water dipole bonds are summarized in Table \ref{tab:h-bonds_and_net-charges}. 
The water O and H atom density distributions and the orientations of the water molecules are compared in Figure \ref{fig:MD_Zhang_0.225_7.00_min}. 
The modified Zhang parameter set generally overestimates the distances between the AD and the water molecules by 0.2 to 0.4 \AA, \  but the relative distance between the O and the H distribution is consistent with the benchmarks (panel a).
The angles of the water dipole moments are slightly overestimated on average but reproduce the orientations of the water molecules quite well, including the slope of $\cos(\alpha)$ in the first hydration shell (panel b). 
We sometimes find molecules with a relatively high $\cos(\alpha)$ value of approx. 0.7 at 4.2~\AA,\ but their number is not statistically significant.
The relationship between the $\alpha$ and $\beta$ angles in the first hydration shell is shown in panel (c). 
In the form of a map, the angles are difficult to compare between MD and DFT, so we group the data into the orientational modes to help us interpret the hydration structure.
Panel (d) shows the probability of a water molecule to be in one of the eight orientational modes.
We see that in the MD-minimized structures it is 25\% less probable to find a water molecule in modes \Romanbar{1} to \Romanbar{3}, compared to the DFT-optimized structures, while modes \Romanbar{5} to \Romanbar{8} have a 25\% higher probability in MD than in DFT. 
However, the qualitative trends from the DFT benchmarks are all maintained upon MD-minimization. 

To further validate the new parameter set, we analyze whether the force field extrapolates to different DDs without loss of accuracy.
We use the new parameter set to minimize DFT-optimized structures of a DI, a TR, and a HE molecule, each surrounded by 160 water molecules. 
For this, we apply the same DFT and MD methods as described in sections \ref{methods:dft_calculations} and \ref{methods:md_simulations} (but with 10 structures per molecule).
In Figure~\ref{fig:MD_Zhang_0.225_7.00_DI_TR_HE_water_density_distributions}, we compare the MD-minimized structures to the DFT-optimized structures in terms of the water O and H density distributions.
As with AD, we see a 0.2 to 0.4 \AA \  shift between the DFT and the MD  results, but the relative distributions between O and H are the same.
Furthermore, in Table \ref{tab:h-bonds_and_net-charges_after_md_min} we compare the MD-minimized structures to the DFT-optimized structures in terms of the number of weak DD-water dipole bonds and the DD net charges. 
In the case of DI and TR, the results are consistent with the AD results. 
As for HE, however, our modified Zhang parameter set overestimates the net charge.
Thus, the analysis shows that the force field trained on AD successfully extrapolates to the larger DDs DI and TR, but further modifications may be needed when attempting to simulate larger structures.

\subsection{MD simulations at room temperature}
The ReaxFF force field with the modified Zhang parameters for the C atoms is now used to investigate the water structure around the DDs which are each hydrated in bulk water at $T=300$~K. 
\begin{figure}
\includegraphics[width=0.48\textwidth]{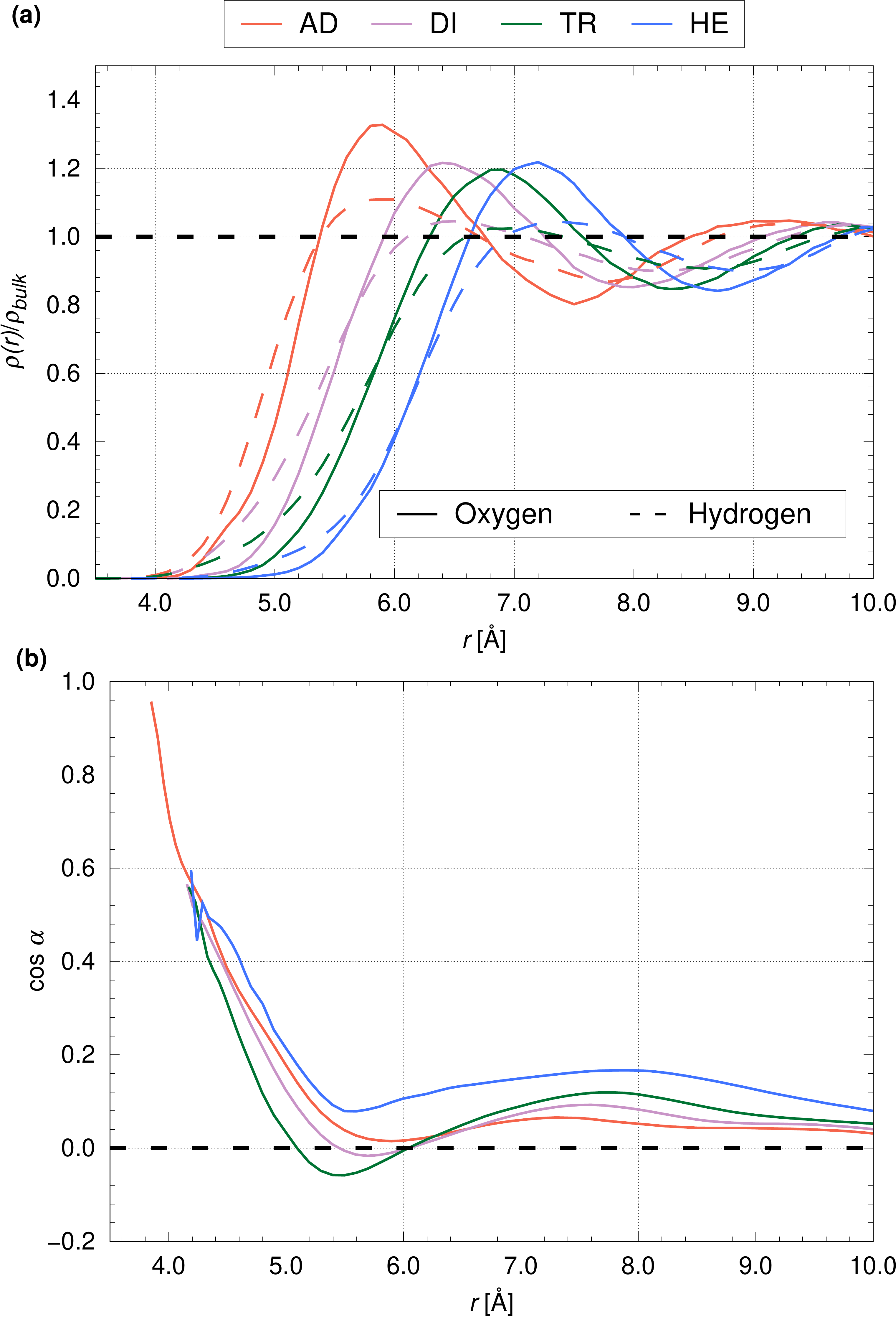}
\caption{ReaxFF simulations of single diamondoids hydrated in bulk water at $T=300$~K and $P=1$~bar using the modified Zhang parameters. 
(a) The radial density distributions of the water O and H atoms as function of the distance to the COM of the DDs, normalized by the bulk density, 0.03344~\AA$^{-3}$ for O and 0.06688~\AA$^{-3}$ for H.
(b) The orientation of the water dipole moments as function of the distance to the COM of the DDs. 
}
\label{fig:MD_Zhang_0.225_7.00_bulk}
\end{figure}
\begin{table}
\begin{tabular}{lcccc}
 & \multicolumn{4}{c}{dipole bonds}\tabularnewline
\cline{2-5}
                & AD & DI & TR & HE \tabularnewline
\cline{2-5}
              DFT 0~K  & 1.3 & 0.2  & 0.6 & 2.3 \tabularnewline   
              MD 0~K   & 1.7 & 0.4  & 1.0 & 1.7 \tabularnewline
              MD 300~K & 0.4 & 0.3 & 0.3 & 0.6 \tabularnewline
\end{tabular}
\, 
\begin{tabular}{cccc}
\multicolumn{4}{c}{DD net charge {[}e{]}}\tabularnewline
\hline 
   AD & DI & TR & HE\tabularnewline
\hline 
-0.183 & -0.133 & -0.13 & -0.21\tabularnewline
-0.197 & -0.251 & -0.127 & 0.303\tabularnewline
-0.298 & -0.297 & -0.185 & 0.188\tabularnewline
\end{tabular}
\caption{The time averaged total number of DD-water dipole bonds (C-H$\cdots$O-H$_2$) and the net charges of the investigated DDs, from the DFT optimization and after MD minimization with the modified Zhang parameters.}
\label{tab:h-bonds_and_net-charges_after_md_min}
\end{table}
Figure~\ref{fig:MD_Zhang_0.225_7.00_bulk} shows the normalized density distributions of the water O and H atoms as function of the distance to the DD COM.
The results reflect typical signatures of hydrophobic solvation of smooth charged spheres, but with quantitative corrections~\cite{Dzubiella2004}.
From AD to TR, the heights of the first peaks decrease monotonically due to increasing attractive forces from the bulk. 
Also, each DDs O and H peaks are located at nearly the same distance from the DD's COM. 
However, the density distributions around the DDs show a weaker structuring compared to the smooth spheres at the same pressure and temperature.
In the case of HE, the height of the first peaks increases again due to a compensating effect of the positive HE net charge (see Table~\ref{tab:h-bonds_and_net-charges_after_md_min}).
The water molecules respond by a cooperative reorientation, as indicated by the H peak shifting slightly away from the position of the O peak.
Further corrections have their origin in the anisotropy of the DD's structures and small-scale interface effects.

Table~\ref{tab:h-bonds_and_net-charges_after_md_min} also shows that for all DDs the number of DD-water dipole bonds further decreases at room temperature compared to the results from MD-minimization.
This seems to indicate that C-H$\cdots$O-H$_2$ dipole bonds are not disturbing the water hydrogen bond network surrounding the DDs.
The few existing dipole bonds may even be a byproduct of the stabilization of the water hydrogen bond network very close to the interface, as we discuss in appendix~\ref{app:dang}. 
\begin{figure}
\includegraphics[width=0.5\textwidth]{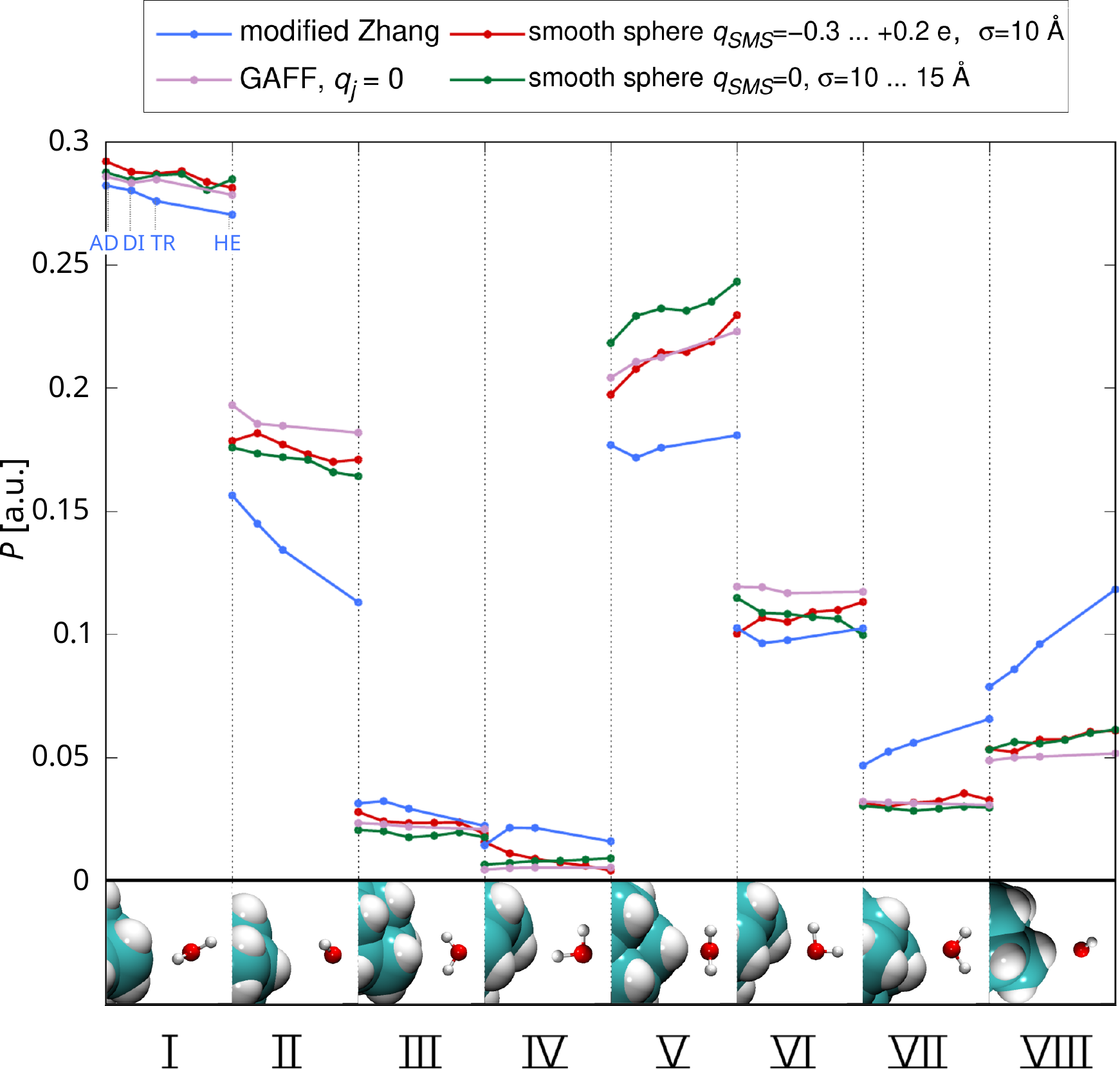}
\caption{
The probability of finding a water molecule in the first hydration shell to be oriented in one of the eight orientational modes.
In the atomistic simulations (blue and purple lines) the dots in each line represent the diamondoids, AD to HE, from left to right. 
In the simulations with GAFF, all partial charges of the DDs were set to zero.
In the simulations with smooth spheres (SMSs) instead of atomistic DDs, the dots represent different net charges $q_{\rm SMS}$ (red lines) and different sizes $\sigma_{\rm SMS}$ (green lines) of the SMSs, all of which correspond to the atomistic DDs. 
All simulations are at $T=300$~K.
}
\label{fig:Structure_HS_smallrange}
\end{figure} 
In Figure \ref{fig:Structure_HS_smallrange} we analyze the water structure of the first hydration shells around the DDs in more detail. 
The figure shows how the increase in molecular size affects the specific orientational modes of the water molecules in the first hydration shell. 
The blue lines show the results for the ReaxFF with the modified Zhang parameters. 
Every column corresponds to an orientational mode and the dots in each line stand for the DD, i.e., AD to HE from left to right. 
Every dot indicates the probability to find a water molecule oriented in the particular mode around the respective DD. 
With increasing DD size we find a strong decrease of the probabilities of modes \Romanbar{1} and \Romanbar{2}, only small changes in modes \Romanbar{3} to \Romanbar{6}  and a strong increase of modes \Romanbar{7} and \Romanbar{8}.
Generally, it becomes less likely for water molecules to point with their H atoms to the DD as the DD size increases.
We suspect  three reasons for this trend. 
First, the change of the DD net charge (see Table~\ref{tab:h-bonds_and_net-charges_after_md_min}). 
Second, the decrease of the DD curvature with increasing DD size.
Third, effects on small length scales at the DD-water interfaces due to the discrete partial charges.

We can isolate the first two contributions from the third one by repeating the simulations with Lennard-Jones smooth spheres (SMS) instead of the DDs.
For this, we replace the DD in our simulations by a Lennard-Jones particle with size $\sigma_{\rm SMS}$ and charge $q_{\rm SMS}$. The mass $m_{\rm SMS}$ is the sum of the DD's constituent atomic masses (e.g. $m_{\rm SMS}=386.529$~amu for AD), and $\varepsilon_{\rm SMS}=0.11$~kcal/mol, which is the same value that is used to simulate aliphatic carbon atoms in the General Amber Force Field (GAFF)~\cite{Wang2004}.
We compare the results from simulations with different $\sigma_{\rm SMS}$ and $q_{\rm SMS}$, to find out which of those properties causes the strongest change of the water orientations in the first hydration shell.
For the simulations with different $\sigma_{\rm SMS}$, the charge of the SMS is set to zero, but the size is varied from $\sigma_{\rm SMS}=10$~\AA\ to $15$~\AA\ in steps of $1$~\AA.
To determine which $\sigma_{\rm SMS}$ corresponds to which DD, we match the density distributions of the water O atoms around the DDs to those around the SMSs. 
We particularly match the position where the density of the rising first peak reaches the bulk value of 0.03344~\AA$^{-3}$. 
For the simulations with different $q_{\rm SMS}$, the size of the hydrated SMSs is the AD size ($\sigma_{\rm SMS}=10$~\AA) and the charges range between $q_{\rm SMS}=-0.3$~e and $+0.2$~e in steps of $0.1$~e, since the DD charges obtained with the modified Zhang parameters range from $-0.2979$~e (AD) to $+0.1874$~e (HE).
To ensure charge neutrality, we compensate the solute charge by adding a counter charge to one randomly selected water molecule.

In Figure \ref{fig:Structure_HS_smallrange} (red and green lines) we review the eight specific orientational modes of the water molecules in the SMS simulations.
In general, the probability to find a water molecule in each mode around a SMS is qualitatively similar to that of the corresponding DD from the atomistic simulation.
Most of the upward and downward trends with increasing solute size also coincide, albeit with much smaller slopes, so that the water molecules are much stronger affected by the atomistic DDs than by the SMSs.
We suspect that this is due to the distribution and strength of the partial charges of the DDs.
To confirm this, we re-simulate the atomistic DDs but this time using GAFF and the TIP3P model for water, with all DD partial charges permanently set to zero.
The response of the water molecules to the uncharged atomistic DDs (purple line in Figure~\ref{fig:Structure_HS_smallrange}) is just as weak as it is to the uncharged SMSs (green line).
Apparently, the water molecules do not distinguish between the atomistic representation of a DD and an equivalently sized SMS, as long as both carry no charge.
Neither do the water molecules respond particularly strongly to the change in the solute size as we see from the uncharged SMSs and from the uncharged GAFF model.
Consequently, it must be the discrete distribution of partial charges in the modified Zhang model which strongly interact with the water molecules on small length scales.

\begin{figure}
\begin{centering}
\includegraphics[width=0.50\textwidth]{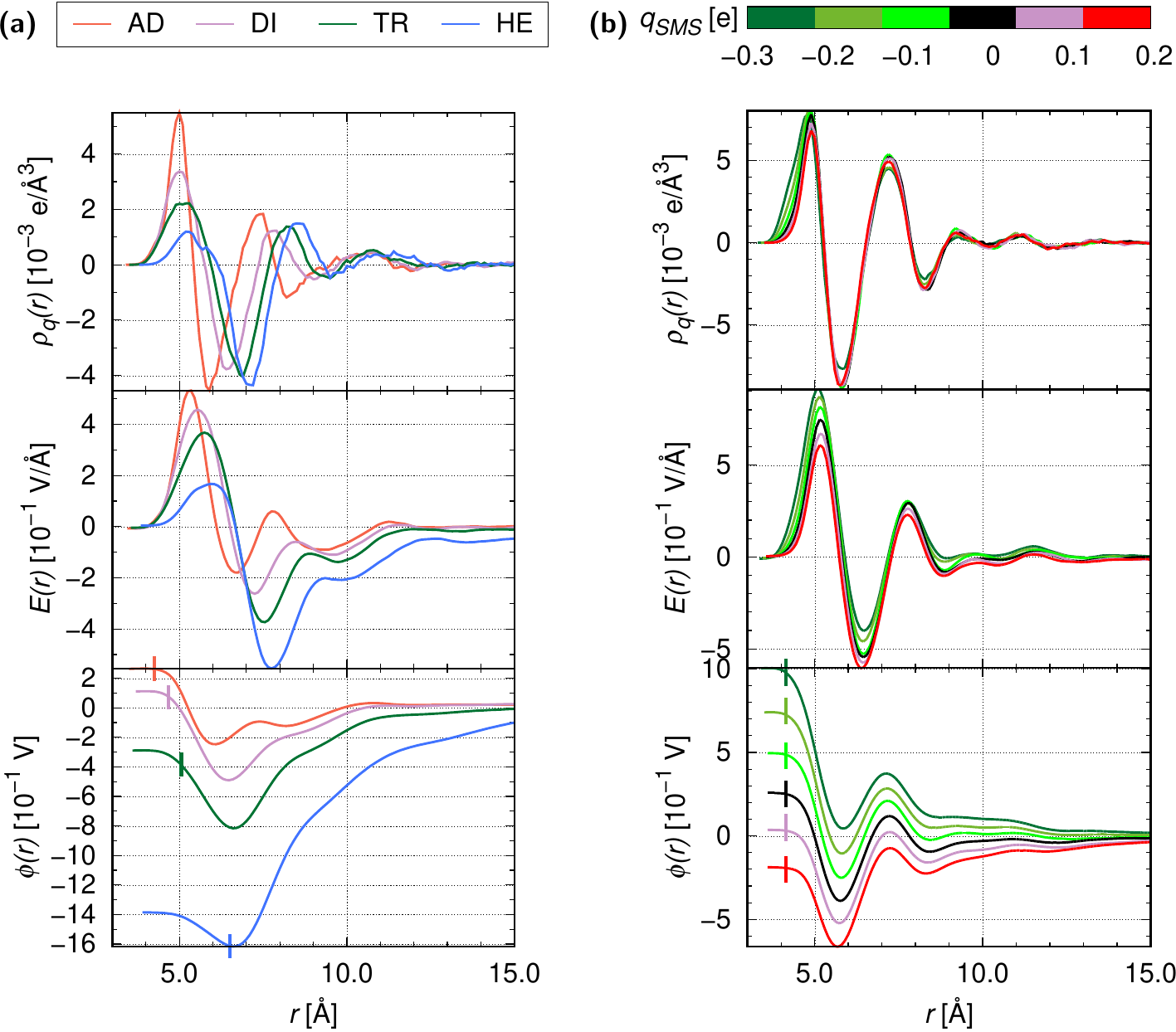}
\par\end{centering}
\centering{}\caption{Electrostatic properties of the bulk water around the hydrated DDs in (a) the ReaxFF simulations with the modified Zhang parameters (net charges are: $-0.2979$~e (AD), $-0.2976$~e (DI), $-0.1852$~e (TR), $+0.1874$~e (HE)) and in (b) the simulations where the DD is replaced by a smooth sphere (SMS) with the charge $q_{\rm SMS}$.
The radial charge density distribution $\rho_{q}(r)$, the radial electric field $E(r)$ and the radial electrostatic potential $\phi(r)$.
The short vertical lines mark the position of the solute surface, which is defined as the Gibbs dividing surface~\cite{Schttl2015}. 
\label{fig:rhoq_e_phi}}
\end{figure}

In the following we show that there is a connection between the changes in the water structure and the electrostatic potential inside the water.
We calculate the radial charge density distributions $\rho_{q}(r)$ of the water with respect to the center of mass of the solutes via 
\[
\rho_{q}(r)=\sum_{j}q_{j}\rho_{j}(r)
\]
and from this the electric field 
\begin{equation}
E(r)=\frac{1}{\epsilon_{r}\epsilon_{0}r^{2}}\int_{0}^{r}\rho_{q}(r^{\prime})r^{\prime2}{\rm d}r^{\prime}
\end{equation}
and electrostatic potential that the water molecules create around the solute
\begin{equation}
\phi(r)=-\int_{0}^{r}E(r^{\prime}){\rm d}r^{\prime}.
\end{equation} 
We do this for the ReaxFF simulations with the modified Zhang parameters and the simulations with the charged SMSs.

The radial distribution of the water charge density around the charged atomistic DDs is shown in Figure~\ref{fig:rhoq_e_phi} (a).
Notably, the charge density is much more sensitive to the solute's net charge at the first positive peak than it is anywhere else, due to the dielectric screening properties of the water.
While the charge of the solute increases, the first peak of the charge density gradually decreases.
This is because the negative net charge of the solute attracts the positively charged H atoms and repels the negatively charged O atoms.
Conversely, if the solute becomes steadily more positive, more H atoms are repelled and O atoms attracted so that the positive peak decreases.
However, the influence of the solute charge on the water molecules is in competition with the intermolecular forces that stabilize the hydrogen bond network at the water interface. 
So, even strongly charged positive solutes will not repel all the H atoms from their water interface layers, which is why the first peak never turns negative in the tested charge range.
With the decrease of the first charge density peak we also see a decrease of the electrostatic potential $\phi$ of the system at the interface compared to the bulk water.
This is because the first hydration layer becomes less and less positive relative to the regions beyond, where the charge density stays almost the same height.
This charge relocation causes a steeper electrostatic potential between the first hydration layer and the bulk water.

In panel (b) we see that the electrostatic properties in the simulations with the charged SMSs qualitatively agree with the charged atomistic DDs.
However, the first peak of the charge density is less sensitive to the solute's net charge and so is, consequently, the electrostatic potential: At the interface it has a total range of 1.7~V in the atomistic simulations and only 1.2~V in the SMS case.
The SMS has a strong influence on the left flank of the first peak but the range of the SMS Coulomb potent-ial seems to be decayed before it reaches the peak.
In contrast, the atomistic DDs exert a longer ranged influence on the water charge distribution, on the one hand because of the DD's different shapes and sizes, and on the other hand because of the more complex interactions between individual DD and water atoms on small length scales.

Lastly, in appendix~\ref{app:dang} we show that these interactions are also responsible for the particular distribution of dangling OH groups in the first hydration shells.

\subsection{Conclusions}
The goals of this work were to create a ReaxFF force field for MD simulations of DDs and to characterize their hydration structure in more detail than has ever been done before.
We reviewed ReaxFF force fields from literature~\cite{Aryanpour2010,Wang2020b,Zhang2018} and realized that they do not yield simulation results that are consistent with our DFT benchmarks.
Thus, we proposed a ReaxFF force field with refitted parameters.

ReaxFF employs polarizable partial charges that describe many-particle interactions on small length scales more accurately than fixed partial charges do.
This allowed us to shed new light on the electrostatic properties of the first hydration shell and its  structure, which has been described as very unusual in experiments~\cite{Petit2017}.
We found a spectrum of water orientations, greatly expanding on what was known before in literature~\cite{Ohisa2011,Doi2013}.
We showed that there are almost no interfacial C-H$\cdots$O-H$_2$ bonds, in which we agree with experiments~\cite{Petit2017}, except at very close distances to the interface, where they contribute to the stabilization of the water-water hydrogen bond network~\cite{https://doi.org/10.48550/arxiv.2102.13312}.
However, we also showed that there is a substantial number of dangling OH groups within the first hydration shell, which have also been observed around NDs and other hydrocarbons~\cite{Mizuno2009,Perera2009,TomlinsonPhillips2011,Davis2012,Davis2013,https://doi.org/10.48550/arxiv.2102.09187}.
Generally, we found that for understanding hydrophobic hydration~\cite{Lum_1999,Chandler_2002} it is of fundamental importance to know the effects of small surface perturbations on hydrophobic behavior.
From the systematic exclusion of atomistic detail we can conclude that the hydration structure is strongly affected by the interactions between the discrete partial charges on small length scales at the DD-water interface.
However, we can also conclude that hydrophobic behavior is qualitatively impervious to small surface perturbations.

Further work is underway to extend the model to simulate the interplay between dopants and electrolytes and to study surface transfer doping and the dynamics of solvated electrons generated from DDs and NDs. 
Our study may also motivate future investigations with a view to a better transferability of ReaxFF force fields to a wider variety of hydrocarbon/water systems.

\section{Acknowledgement}
The authors acknowledge support by the state of Baden-Württemberg through bwHPC, and support from the high performance computing cluster Curta of Freie Universität Berlin. 
TK, JD, and AB acknowledge the support of the Helmholtz Einstein International Berlin Research School in Data Science (HEIBRiDS).
The authors wish to thank Dr. Sébastien Groh for his help on ReaxFF and inspiring discussions.

\appendix
\appendixpage
\addappheadtotoc

\section{Comparison to existing ReaxFF force fields}
\label{appA}

\begin{figure*}
\includegraphics[width=1\textwidth]{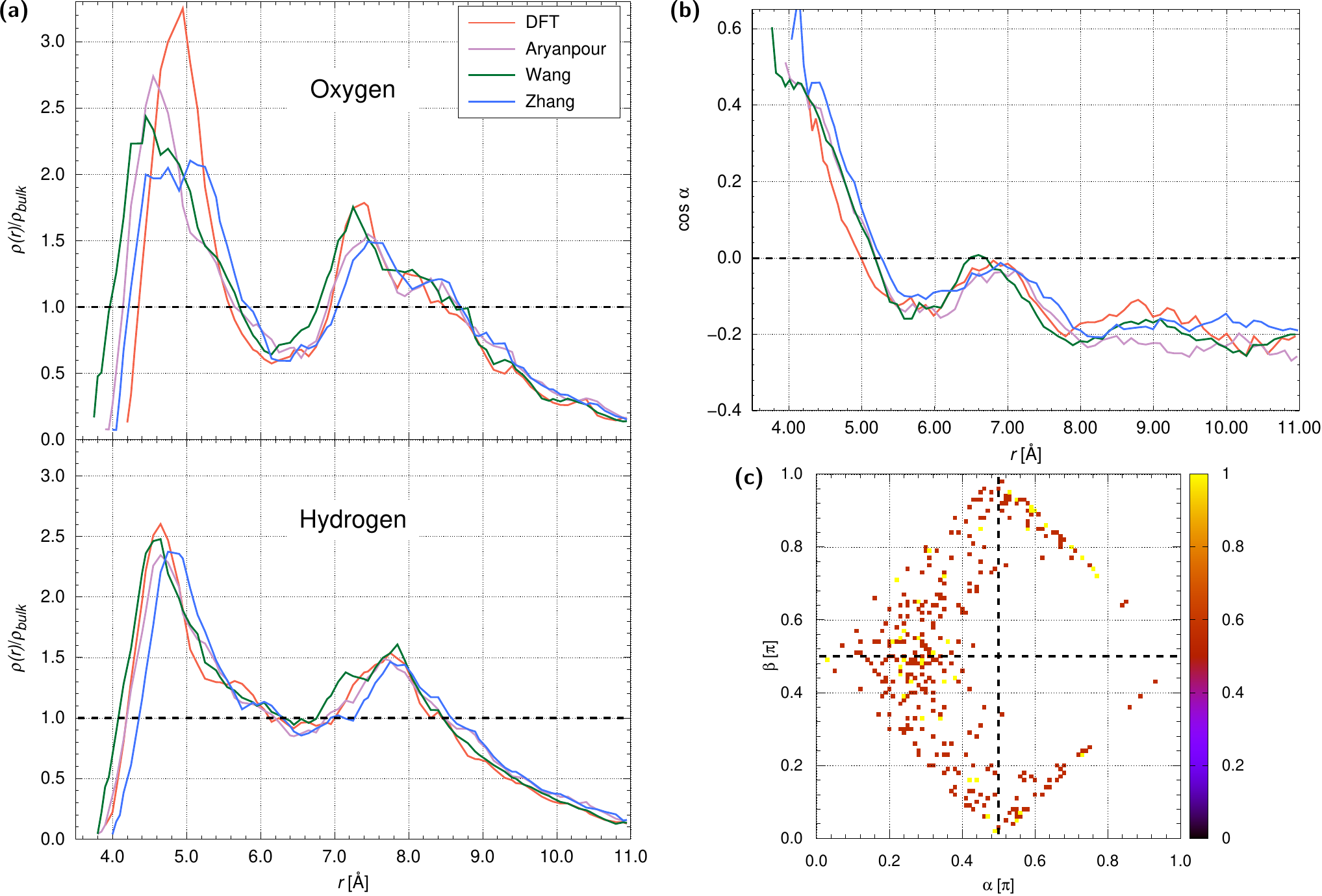}
\caption{
Results of the energy minimization using different ReaxFF force fields and comparison with the DFT benchmarks.
a) The radial density of the water O and H atoms as function of the distance to the AD COM, normalized by the bulk density, 0.03344~\AA$^{-3}$ for O and 0.06688~\AA$^{-3}$ for H. 
b) The orientation of the water dipole moments.
c) The relationship between the angles $\alpha$ and $\beta$ in the first hydration shell obtained with the Zhang parameters.
}
\label{fig:MD_min_Reaxff}
\end{figure*}

After the five quantities, that will serve as our benchmarks for the force field fitting, have been established from the DFT-optimized structures in section~\ref{sec:DFT_calculations}, we minimize those structures using the ReaxFF force fields with the Aryanpour, Wang and Zhang parameter sets. 
In the following, we analyze the differences between the MD-minimized structures and the DFT-optimized structures.  
The differences are presented in Figure \ref{fig:MD_min_Reaxff} and Table \ref{tab:h-bonds_and_net-charges}. 
Panel (a) compares the density distributions for the water O and H atoms in the first hydration shell. 
We pointed out that in the DFT-optimized structures the space closest to the AD is dominantly populated by H atoms, while the O atoms are located approx. $0.3$~\AA \ further away. 
With that in mind, panel (a) exposes the biggest differences between the MD-minimized and the DFT-optimized structures:
In all three ReaxFF results, the O atoms are located as close as or closer to the AD than the H atoms. 
In the DFT structures, there are no O atoms at $r < 4.2$~\AA, but in the MD structures, we find a significant number of O atoms at this position.
This is reflected in the orientations of the dipole moment vectors in Figure \ref{fig:MD_min_Reaxff} (b).
At $r\geq4.2$ \AA, \  all the values from the ReaxFF are in good agreement with the DFT results. 
At $r < 4.2$~\AA, we find O atoms in the MD structures with the average $\cos\left(\alpha\right)$ rising higher, the closer the O atom is to the AD. 
This result means that there are more water molecules with their OH groups pointing away from the AD than in the DFT structures.  
This corresponds to the $\alpha$-$\beta$ map in panel (c), which describes the water orientations resulting from the Zhang parameters.
Comparing to Figure \ref{fig:DFT_results} (c) we see that the orientation modes \Romanbar{7} and \Romanbar{8} are much more prevalent after the MD-minimization.
These are all molecules that are pointing with both H atoms away from the AD, due to the electrostatic repulsion between the positive water H atoms and the net-positive AD molecule.

\begin{table}
\begin{tabular}{lccccc}
 & \multicolumn{5}{c}{number of C-H$\cdots$O-H$_2$ dipole bonds}\tabularnewline
\cline{2-6} \cline{3-6} \cline{4-6} \cline{5-6} \cline{6-6}  
                & DFT & Aryanpour & Wang & Zhang & modified Zhang\tabularnewline
\cline{2-6} \cline{3-6} \cline{4-6} \cline{5-6} \cline{6-6}    
            0~K  & 1.3 & 5.4 & 5.2 & 4.5 & 1.7\tabularnewline   
          300~K  & n/a & 5.4 & 14.7 & 4.5 & 0.4\tabularnewline
\tabularnewline
 & \multicolumn{5}{c}{AD net charge {[}e{]}}\tabularnewline
\cline{2-6} \cline{3-6} \cline{4-6} \cline{5-6} \cline{6-6}  
                & DFT & Aryanpour & Wang & Zhang & modified Zhang\tabularnewline
\cline{2-6} \cline{3-6} \cline{4-6} \cline{5-6} \cline{6-6}   
            0~K  & -0.183 & 1.217 & 1.397 & 0.896 & -0.197\tabularnewline
          300~K  &    n/a & 1.350 & 2.037 & 1.138 & -0.298\tabularnewline
\end{tabular}
\caption{The time averaged total number of dipole bonds between the AD and water (C-H$\cdots$O-H$_2$) and the net charge of the AD. Our results from DFT (water droplet at 0~K) and ReaxFF with the modified Zhang parameters (bulk water at 300~K) are compared to results obtained with the Aryanpour~\cite{Aryanpour2010}, Wang~\cite{Wang2020b}, and Zhang~\cite{Zhang2018} ReaxFF parameters.}
\label{tab:h-bonds_and_net-charges}
\end{table}

In MD, the net charge of the AD is defined as the sum of all atomic partial charges of the AD molecule.
With the tested ReaxFF parameters, the QEq method yields a net charge of approx. $+1$~e for the AD, averaged over all structures, while the net charge from DFT was $-0.183$~e (table \ref{tab:h-bonds_and_net-charges}). 
Due to the positive net charge the polar water molecules are strongly directed by the electrostatic field of the AD.
As a consequence, the O atoms become potential acceptors for C-H$\cdots$O-H$_2$ dipole bonds. 
The number of dipole bonds ranges from 4.5 to 5.4 in the MD-minimized structures which is much larger compared to 1.3 in the DFT case.

To conclude this section, we find that the commonly used ReaxFF parameters for describing hydrocarbon-water interactions do not yield a good agreement with the DFT-optimized water distribution around the AD. 
The net charge of the AD is very different and as a result the orientations and positions of the water molecules in respect to the AD COM are not correctly reproduced.
The number of dipole bonds between the AD and the water molecules is overestimated by a factor of approx. 4. 
Thus, to fit the ReaxFF simulations to the DFT-optimized structures, it is necessary to modify one of the existing parameter sets. 

\section{Fitting of the force field parameters}
\label{appB}

\begin{table}
\begin{tabular}{ll}
benchmark quantities $x_{i}$ & weight $\sigma_{i}$\tabularnewline
\hline 
O profile & 0.01 \AA  $^{-3}$\tabularnewline
H profile & 0.02 \AA  $^{-3}$\tabularnewline
$\cos{\alpha}$  & 0.1\tabularnewline
$\alpha,\beta$-pairs & 0.1\tabularnewline
number of C-H$\cdots$O-H$_2$ bonds & 0.13\tabularnewline
net charge & 0.0183 e\tabularnewline
\hline 
\end{tabular}
\caption{The weighting coefficients for the cost function (eq.~\ref{eq:cost_function}) }
\label{tab:tolerances}
\end{table}

\begin{figure}
\includegraphics[width=0.4\textwidth]{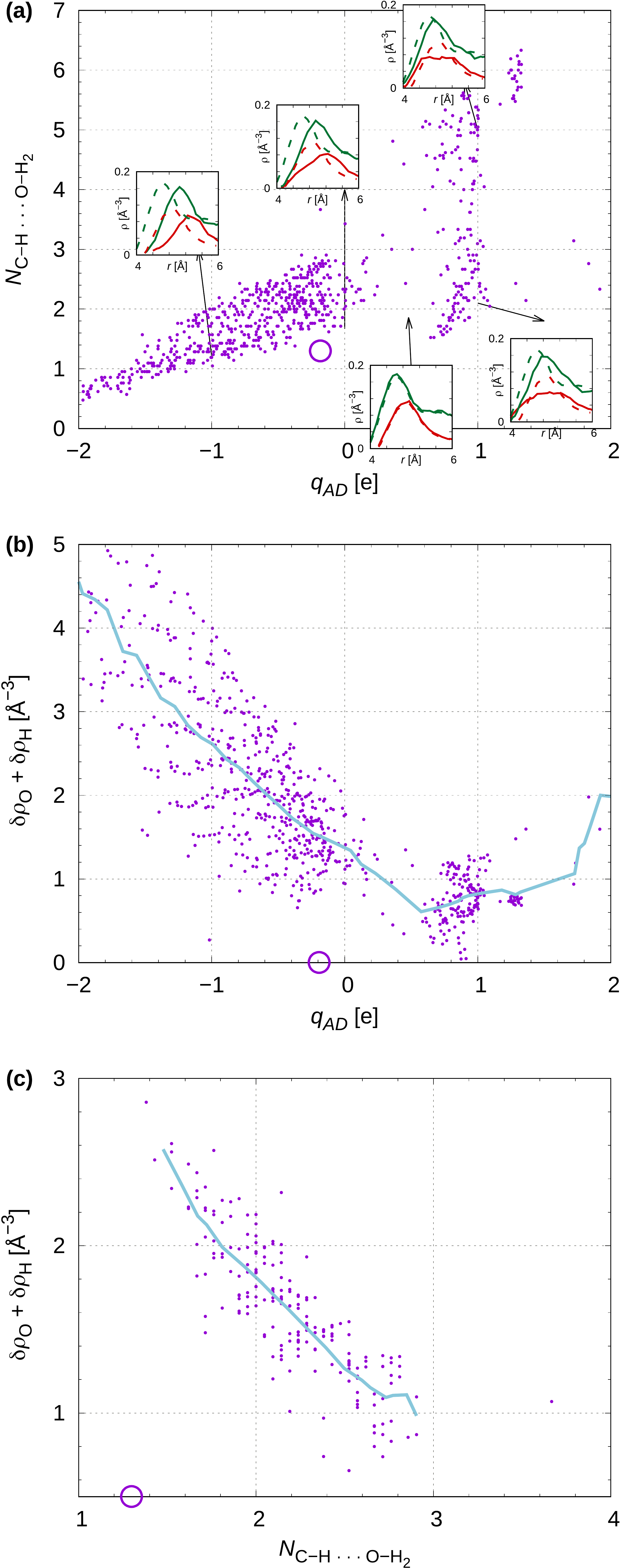}
\caption{
Trends from the QEq parameter study.
Each point represents a simulation with one distinct combination of the QEq parameters. 
Only those simulations are shown in which the AD molecule did not deform during the minimization. 
a) The relation between the charge and the number of weak AD-water dipole bonds. 
The purple circle marks the target values for the optimal fit. 
However, no parameter set yields the desired combination of results.
The insets show the O (red) and H (green) radial density distributions in the first hydration shell.
The dashed lines are the references from the DFT-optimization and the solid lines are the ReaxFF results.
b) The relation between the charge and the deviation of the water distribution from the DFT benchmark. 
The blue line is the running average.
c) The relation between the number of dipole bonds and the deviation of the water distribution. 
Only simulations in which the net charge is between $-0.5$ and $0$~e are represented in this panel.
}
\label{fig:parameter-search-results}
\end{figure}

Since none of the tested parameter sets stands out positively in our comparison between MD and DFT, we choose the Zhang set for modification, because the H and O parameters are especially optimized to describe structural and dynamic properties of liquid water.
For our modifications we assume that the overemphasis of the O atom density close to the AD is due to the electrostatic interactions between the AD and the water molecules. 
The positive net charge of the AD likely exerts an attractive force on the O atoms and a repulsive force on the H atoms. 
As we do not want to change the parameters of the H and O atoms, which would lower the high quality of the water-water interactions, we can tune the C atom parameters associated with the QEq method to bring the net charge down to the benchmark value.
Additionally, the increase of the O atom density can also be explained by the emerging weak AD-water dipole bonds.
Tuning the QEq parameters of the C atoms has the biggest effect on the donor strength and therefore also on the number of dipole bonds and the O and H density distributions.
As described in the methods section \ref{sec:fitting}, we systematically modify the shielding $\gamma$, electronegativity $\chi$ and hardness $\eta$ of the C atoms. 
The desired combination of these three parameters minimizes the cost function equation~\ref{eq:cost_function}. 
The weighting coefficients are listed in Table~\ref{tab:tolerances}.

Although the parameter study is of a rather technical nature, it is worth to report on it for future reference.
We observe a few interesting trends. 
First, we see a dependence of the weak AD-water dipole bonds on the AD net charge in Figure~\ref{fig:parameter-search-results} (a).
The purple circle marks the target charge and dipole bond count from the DFT benchmark structures.
Within the investigated range, no QEq parameter combination exactly reproduces the target values from the DFT-optimization, so we have to accept compromises.
There is also a noticeable gap in panel (a) between $q_{AD}=0$ and $q_{AD}=1$~e which only a few QEq parameter combinations were able to fill.
Ironically, the best fits of the density distributions can be found within this gap.
One of the compromises is that the water molecules will have a higher distance from the AD COM than they have in the DFT results.
Figure~\ref{fig:parameter-search-results} (b) shows that on average the best match of the water O and H atom density distributions is achieved, if the AD net charge is between $+0.5$~e and $+1$~e. 
However, the target value for the charge is $-0.183$e (circle).
Figure~\ref{fig:parameter-search-results} (c) shows that the water atom density distributions match the DFT-optimized structures better if the number of weak AD-water dipole bonds is higher. The target value, however, is 1.3 (circle). 
Ultimately, the cost function (equation~\ref{eq:cost_function}) is minimized by QEq parameter values that compromise on all benchmark quantities (see section \ref{sec:Benchmarking_results}).

\section{Dangling OH groups in the first hydration shell}
\label{app:dang}
\begin{figure}
\includegraphics[width=0.47\textwidth]{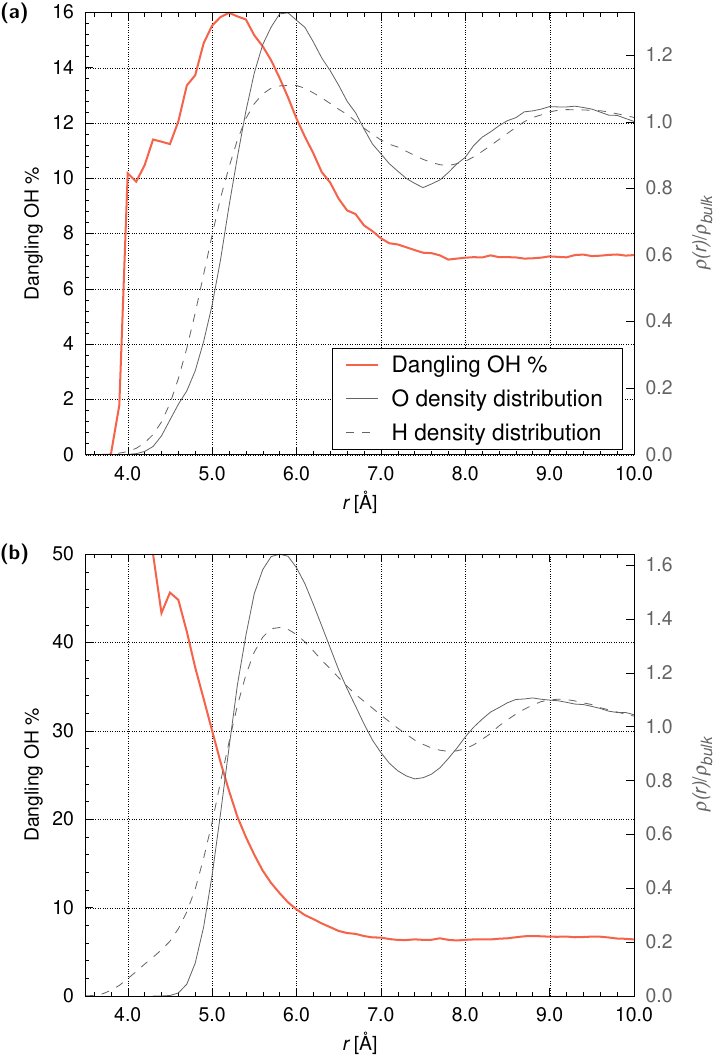}
\caption{The average fraction of non-hydrogen bonded OH groups (i.e. dangling OH bonds) per water molecule around the AD at $T=300$~K in the simulations with (a) the modified Zhang parameters and (b) the SMS with $q_{\rm SMS}=-0.3$~e and $\sigma_{\rm SMS}=10$~\AA. 
The grey curves show the radial density distributions of the water O and H atoms.
}
\label{fig:dangling_OH}
\end{figure}
As described in the introduction, experiments on small hydrogenated NDs suggest that their first hydration shells contain water molecules with dangling OH groups, which is an often observed phenomenon for nonpolar (hydrocarbon) solutes. 
From our trajectories at 300~K in bulk water, we calculate the average number of OH groups per water molecule within a distance $r$ from each solute's COM that are not engaged in any hydrogen bond. 
Figure \ref{fig:dangling_OH} (a) shows the result obtained from our ReaxFF simulation with the modified Zhang parameter set. In panel (b), the same is shown from the simulation of a hydrated SMS with the parameters $q_{\rm SMS}=-0.3$~e and $\sigma_{\rm SMS}=10$~\AA.
For better overview, we juxtapose the fractions of non-hydrogen bonded OH groups with the water O and H atom density distributions from the same simulations (gray lines). 
In both simulations every water molecule outside of the first hydration shell, donates, on average, 1.85 hydrogen bonds.
This is already known from literature~\cite{Laage2006}.
Conversely, at any given time, 7.6\% of all OH groups are not engaged in hydrogen bonding in the bulk.
They are most likely in the transition state between two hydrogen bonded structures~\cite{TomlinsonPhillips2011}.
Closer to the solute, in both simulations, the number of non hydrogen bonded OH groups increases in the first hydration shell. 
In the ReaxFF simulation, the ratio decreases again after reaching a peak value of 16\% at $r=5.2$~\AA. 
This is relatively high compared to measurements around small hydrocarbons such as neopentane for instance~\cite{TomlinsonPhillips2011}, but it has been shown that the fraction of dangling OH groups increases with the solute size~\cite{Perera2009}.
In the SMS case, it increases to 50\% until it reaches the interface. 
The difference between the models can be related to the orientational modes shown in Figure~\ref{fig:Structure_HS_smallrange}.
In the SMS simulations, the water H atoms are attracted by the negative net charge, which destabilizes the hydrogen bond network and stabilizes the dangling OH groups. 
This is consistent with experimental studies of dangling OH groups around negatively charged hydrocarbons~\cite{Davis2013}.
The same happens in the ReaxFF simulations, but at distances more than 1~\AA\ away from the AD.
Closer to the AD however, some water H atoms are repelled by the positive partial charges of the AD H atoms - a mechanism that has been seen in MD simulations of H-terminated NDs with zero net charge~\cite{https://doi.org/10.48550/arxiv.2102.13312}. 
This is the reason for the elevated number of water molecules with the orientational modes \Romanbar{7} and \Romanbar{8}, in which the OH bonds are both oriented away from the AD towards the water. 
This stabilizes the hydrogen bond network for water molecules very close to the AD.
It also means that water molecules can reside so close to the AD only if they are part of the water hydrogen bond network. This is probably the reason why we observe a small amount AD-water dipole bonds in our simulations.
\clearpage


\providecommand{\latin}[1]{#1}
\makeatletter
\providecommand{\doi}
  {\begingroup\let\do\@makeother\dospecials
  \catcode`\{=1 \catcode`\}=2 \doi@aux}
\providecommand{\doi@aux}[1]{\endgroup\texttt{#1}}
\makeatother
\providecommand*\mcitethebibliography{\thebibliography}
\csname @ifundefined\endcsname{endmcitethebibliography}
  {\let\endmcitethebibliography\endthebibliography}{}

\end{document}